\documentclass[%
 reprint,
footinbib, 
 amsmath,amssyacsmb,
 aps,
 prr,
]{revtex4-1}

\usepackage{comment} 
\usepackage[dvipdfmx]{graphicx}
\usepackage{dcolumn}
\usepackage{bm}

\usepackage{float}
\usepackage{ulem}

\usepackage{color}

\begin{document}

\title{
Spatio-temporal chaos of one-dimensional thin elastic layer with the rate-and-state friction law
}
	\author{Yutaka Sumino}
	\email{ysumino@rs.tus.ac.jp}
	\affiliation{%
		Department of Applied Physics, Faculty of Science Division I, Tokyo University of Science, 6-3-1 Nijuku, Katsushika, Tokyo 125-8585, Japan}
        \affiliation{WaTUS and Division of colloid and interface science, Research Institute for Science \& Technology, Tokyo University of Science, 6-3-1 Nijuku, Katsushika, Tokyo 125-8585, Japan
	}
	\author{Takuya Saito}
	\email{tsaito@phys.aoyama.ac.jp}
	\affiliation{%
	    Department of Physical Sciences, Aoyama Gakuin University, 
		5-10-1 Fuchinobe, Chuo-ku, Sagamihara-shi, Kanagawa 252-5258, Japan
	}
	
	\author{Takahiro Hatano}
	\affiliation{%
		Department of Earth and Space Science, Osaka University, 1 Machikaneyama, Toyonaka, Osaka 560-0043, Japan
	}
	
	\author{Tetsuo Yamaguchi}
	\affiliation{%
	Department of Biomaterial Sciences, The University of Tokyo,  
1-1-1 Yayoi, Bunkyo-ku, Tokyo 113-8657, Japan}
	
	\author{Satoshi Ide}
\affiliation{%
	Department of Earth and Planetary Science, Graduate School of Science, The University of Tokyo, 7-3-1 Hongo, Bunkyo-ku, Tokyo 113-0033, Japan}

\date{\today}
	\begin{abstract}
	    {\color{black}
	    Independent of specific local features, global spatio-temporal structures in diverse phenomena around bifurcation points are described by the complex Ginzburg-Landau equation (CGLE) derived using the reductive perturbation method, which includes prediction of spatio-temporal chaos.
    The generality in the CGLE scheme includes oscillatory instability in slip behavior between stable and unstable regimes.
	    Such slip transitions accompanying spatio-temporal chaos is expected for frictional interfaces of a thin elastic layer made of soft solids, such as rubber or gel, where especially chaotic behavior may be easily discovered due to their compliance.
	    Slow earthquakes observed in the aseismic-to-seismogenic transition zone along a subducting plate
	    are also potential candidates.
	    This article focuses on the common properties of slip oscillatory instability from the viewpoint of a CGLE approach by introducing a drastically simplified model of an elastic body with a thin layer, whose local expression in space and time allows us 
	    to employ conventional reduction methods. 
	    Special attention is paid to incorporate a rate-and-state friction law supported by microscopic mechanisms beyond the Coulomb friction law. 
	   We discuss similarities and discrepancies in the oscillatory instability observed or predicted in soft matter or a slow earthquake.
	    }
	\end{abstract}

\maketitle


\section{Introduction}
{\color{black}
Slip instability
is ubiquitously found in a rich variety of 
phenomena~\cite{Strogatz2001,Rice1983}.
Indeed, we quite often encounter examples in daily life.
For instance, the fascinating tones created by musical string instruments such as violins arise from stick-slip motion~\cite{Fletcher1998}.
Another example is vehicles equipped with tires, 
where suppression of oscillatory instability is necessary to secure safety~\cite{Persson1997}.
The last important example given here is earthquakes~\cite{Marone1988, Scholz2002}.
Megathrust earthquakes occur in locked faults on subducting tectonic plates
by suddenly releasing elastic energy stored by plate motion.
Such slip instability emerges at the interface in the elastic body, and thus it is necessary to focus on the interfacial balance between friction and elasticity, in order to gain insights into the underlying physics.

Unstable slip motion may be generally involved in highly or weakly nonlinear components.
This article focuses on weak nonlinearity around bifurcation points.
By doing so,
we construct a complex Ginzburg-Landau equation (CGLE)~\cite{Kuramoto1984,Aranson2002,Sugiura2014} that
extracts common global structures without relying on specific microscopic features. 
It is worth noting that the CGLE offers a generic methodology applicable not only to slip instability, but also to instabilities near bifurcation points, such as transitions found in superconductivity, liquid crystals, Bernard convection, and Taylor vortices~\cite{Kuramoto1984,Aranson2002}. 
The general framework of the CGLE is derived through the reductive perturbation method, which
provides a simple description with a few degrees of freedom and parameters by eliminating fast variables in the original governing equations.

{\color{black}Our main interest is chaotic behavior manifesting in the weakly nonlinear slip instability around bifurcation points.
This is one of the possible phenomena predicted by the CGLE framework,
but the relevance of the chaos to slip instability at an interface of elastic media is still elusive.}
Systems made from rubber and gel would become feasible candidates exhibiting spatio-temporal chaos in laboratory experiments if an appropriate soft elastic solid is chosen from elastics with a broad range of compliance~\cite{Baumberger2006, Baumberger2002, Yamamoto2014, Maegawa2016, Ben-David2010}. 
Indeed, flexibility is a key parameter because distinctly heterogeneous stick-slip motions have been observed between a hard PMMA block and soft PDMS gels~\cite{Yamaguchi2011}.
Recall here the rate-and-state friction (RSF) law discovered in rocks.
{\color{black}The friction law acting on an interface even of a soft elastic solid is quite often of the RSF law family.
Therefore, not only soft solids like gels, but also rigid materials such as rocks of importance to geoscience, are within the scope of the RSF law~\cite{Baumberger2006}.}
In fact, this study was inspired by slow earthquakes observed in subduction zones~\cite{Obara2002,Obara2016}.
While regular or megathrust earthquakes accompany the high nonlinearity very far from the bifurcation point~\cite{Marone1988}, slow earthquakes could have common properties described by weakly nonlinear analyses because they occur in the transition zone between a locked and continuously creeping fault, which is reminiscent of the system parameter varying across the bifurcation point. 

\begin{figure}
    \centering
    \includegraphics{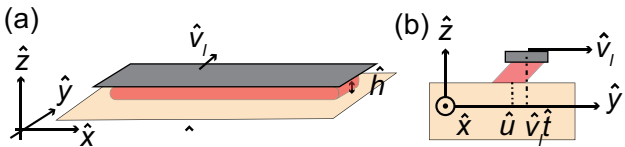}
    \caption{Schematic representations of elastic thin layer model.
    (a) {\color{black}Three-dimensional view.}
    (b) Magnified cross-sectional view {\color{black}on the $\hat{y}$-$\hat{z}$ plane of the coordinate system fixed in the lower plate.
    The upper plate is dragged with velocity $\hat{v}_l$ along the positive direction of the $\hat{y}$-axis, and
    displacements $\hat{u}(\hat{x},\hat{t})$} directed to $\hat{y}$-axis depend on $\hat{x}$.
    }
        \label{thin_layer}
\end{figure}

This paper discusses our analytical and numerical findings {\color{black}in the context of} the CGLE based on weakly nonlinear analyses of a thin layer elastic body with the RSF law.
The analyses are constructed through {\color{black}the reductive perturbation method~\cite{Kuramoto1984} that simplifies the original governing equation.}
This simplification is more than an approximation because it finds the universal features that become manifest through the elimination of specific local properties.
In sec.~\ref{ContinuumBody}, we first introduce a thin elastic layer model under the RSF
law. 
In sec.~\ref{sec:wna}, linear and weakly non-linear analyses are applied near the bifurcation point to show that the slow and global structures are represented by the CGLE, which may display spatio-temporal chaos due to Benjamin-Feir (BF) instability.
Section~\ref{Numerical} demonstrates a numerical simulation, {\color{black} where time evolution is obeyed by} the original governing equation before the reductive perturbation method, and 
{\color{black}then compares the results with the analytical calculations.}
An interesting point is that
the size distribution of slip events numerically obtained for the chaotic regime is exponential, invoking the cumulative distribution of the seismic energy rate reported for slow earthquakes~\cite{Yabe2014}.
In sec.~\ref{Discussion}, we consider the results of soft material experiments and those for slow earthquakes in light of the CGLE.
Section~\ref{Conclusion} concludes this study.
}

\section{Thin layer model with the RSF law}
\label{ContinuumBody}
{\color{black}
We begin with an introduction to an elastic thin layer system.
A flat plate or sheet with thickness $\hat{h}$ is placed on an $\hat{x}$-$\hat{y}$ plane, and the $\hat{z}$-axis points upward so as to meet with the right-handed coordinate [see Fig.~\ref{thin_layer}\,(a)]. 
Friction acts on the bottom interface.
The top boundary is dragged with constant loading velocity $\hat{v}_l$ along the $\hat{y}$-axis, as shown in Fig.~\ref{thin_layer}\,(b).
The layer is sheared and displaced along the $\hat{y}$-axis, to which the $\hat{x}$-axis is perpendicular.
That is, we consider mode III, where displacements denoted by $\hat{u}(\hat{x},\hat{t})$ are uniform for the $\hat{y}$-component, so that we can deal with the system as one-dimensional and spanning the $\hat{x}$-direction.}

{\color{black}
Let us now move on to the governing equation. The elasticity is expressed by Navier's equation.}
When the thickness of the elastic layer $\hat{h}$ is small enough, imposing the boundary condition on Navier's equation simplifies the equation of motion for the thin layer {\color{black}(see Appendix F)}
as

\begin{align}\label{eqn02}
\hat{\rho}\hat{h}\frac{\partial^2 \hat{u}(\hat{x}, \hat{t}) }{\partial \hat{t}^2}=&\hat{G}\hat{h}\frac{\partial^2 \hat{u}(\hat{x}, \hat{t}) }{\partial \hat{x}^2} 
+\frac{2\hat{G}}{\hat{h}} \left[\hat{v}_l \hat{t} - \hat{u}(\hat{x},\hat{t}) \right]
\notag \\ 
&
-\mu(\hat{x},\hat{t}) \hat{\sigma}\, \mathrm{sgn}\left(\hat{v} \right),
\end{align}
where $\hat{t}$, $\hat{\rho}$, $\hat{G}$, $\mu$, $\hat{\sigma}$, and $\hat{v}\equiv d\hat{u}/d{\hat{t}}$ denote the time, mass density, shear modulus for the elastic body, friction coefficient, normal stress, and velocity along the $\hat{y}$-axis, respectively.
On the right side, the first two terms for the elastic force are derived from the continuum body model for the thin layer, where the interaction range is finite and is scaled by the thickness $\simeq\hat{h}$.
The restoring force for the global inhomogeneous deformation along the $\hat{x}$-axis is replaced with the operator ($\partial^2 /\partial \hat{x}^2$), which serves as the local coupling.
In addition, the top boundary condition emerges as loading in the second term.
The last term is the friction that obeys the RSF law~\cite{Kawamura2012, Ranjith1999a, Rathbun2013}, which acts on the bottom boundary interface with magnitude $\mu (\hat{x}, \hat{t}) \hat{\sigma} $.
The RSF law is commonly exploited to describe rock friction~\cite{Morrow2017, Marone1998a}, and is also known to be readily applicable to other materials, as seen in the literature~\cite{Heslot1994, Baumberger2006}.  
In addition, it is often used to describe regular~\cite{Scholz1998} and slow~\cite{Shibazaki2012} earthquakes. {\color{black} Further, the RSF law friction model {\color{black}has been known to serve as the premise of slip instability related to} nonlinear dynamics, including chaotic dynamics~\cite{Viesca2016a, Viesca2016b, Ranjith2014, Brener2018}.}
The constitutive law for the RSF is written as
\begin{equation}\label{constitutive_law}
\mu = \mu_{*} + a\ln\left(\frac{\hat{v}}{\hat{v}_{*} }\right) +b\ln\left(\frac{\hat{\theta}}{\hat{\theta}_{*} }\right),
\end{equation}
where $\hat{v}_*\equiv \hat{D}_c/\hat{\theta}_*$, and $\hat{\theta}$ denotes a variable that 
{\color{black} represents the interface state depending on the slip history.}
As an evolution law, this paper adopts the slip law~\cite{Rice1983, Ranjith1999},
\begin{equation}\label{evolution_law}
\frac{d \hat{\theta}}{d\hat{t}} = -\frac{\hat{v}\hat{\theta}}{\hat{D}_c} \ln{\frac{\hat{v}\hat{\theta}}{\hat{D}_c}},
\end{equation}
because it provides a better description of a fairly symmetric response with the characteristic distance $\hat{D}_c$ to a
discontinuous increase or decrease in stress
~\cite{Ampuero2008}.
The RSF law gives the steady-state friction as
\begin{equation}
\mu=\mu_{*} + (a-b)\ln\left(\frac{\hat{v}}{\hat{v}_{*} }\right),
\end{equation}
meaning that velocity weakening occurs for $a-b<0$, while velocity strengthening occurs for $a-b>0$.
A Hopf bifurcation that induces unstable slip behavior is observed below around $a-b\simeq 0$ and $a-b<0$, which are the main regions of interest for slow earthquakes.

In preparation for the reductive perturbation method,
we recast Eqs.~\eqref{eqn02}-\eqref{evolution_law} as time evolution Eqs.~\eqref{xi_time_evo}-\eqref{rs07}: 
\begin{eqnarray}\label{xi_time_evo}
\frac{d}{d t} 
\left(
\begin{array}{c}
\xi_1\\
\xi_2\\
\xi_3
\end{array}
\right)
=
\left(
\begin{array}{c}
F_1(\bm{\xi})\\
F_2(\bm{\xi}) +
c^2 \frac{\partial^2 \xi_1}{\partial x^2} \\
F_3(\bm{\xi})
\end{array}
\right)\label{eq-rsall}
\end{eqnarray}
where
\begin{eqnarray}
F_1(\bm{\xi}) &=& \xi_2,
\label{rs05}
\\
F_2(\bm{\xi}) &=& - G \xi_1
\label{rs06}\\
&& - \left[ \frac{\mu_{*}}{a} + \xi_3 +\ln\left(\frac{\xi_2 +v_l}{v_{*} }\right)\right]\mathrm{sgn}\left(\xi_2+v_l \right),
\nonumber
\\
F_3(\bm{\xi}) &=& -(\xi_2+v_l) \left[\xi_3 + (1+\beta) \ln\left(\frac{\xi_2+v_l}{v_*}\right) \right],
\label{rs07} \end{eqnarray}
with three dimensionless variables:
\begin{align}
& &\xi_1=u-v_l t,& &\xi_2=v-v_l,& & \xi_3=(1+\beta)\ln{\frac{\theta}{\theta_*}}.
\end{align}
Note that, for compact notation, the units for length, time, and stress are introduced, respectively, as follows:
	\begin{eqnarray}\label{Eq_unit}
	\hat{D}_c, \qquad \hat{t}_u\equiv \sqrt{\frac{\hat{D}_c \hat{\rho} \hat{h}}{a\hat{\sigma}}}, \qquad a\hat{\sigma},
	\end{eqnarray}
	We use the dimensionless variables
	$t=\hat{t}/\hat{t}_u$, $x=\hat{x}/\hat{D}_c$,$u=\hat{u}/\hat{D}_c$, $h=\hat{h}/\hat{D}_c$, $v=\hat{v}\hat{t}_u/\hat{D}_c$, $v_l=\hat{v}_l\hat{t}_u/\hat{D}_c$, $v_*=\hat{v}_*\hat{t}_u/\hat{D}_c$.
In addition, parameters associated with the physical properties are listed below:
\begin{eqnarray}
G
=\frac{2\hat{G} \hat{D}_c}{a\hat{\sigma} \hat{h}},
\quad
\beta=\frac{b}{a}-1,
\quad
c = \frac{\hat{t}_u}{\hat{D}_c} \sqrt{\frac{\hat{G}}{\hat{\rho}}},
\end{eqnarray}
where $G$ represents the coupling with the driving factor provided by the upper plate, $\beta$ is the distance from the velocity weakening point ($a-b=0$), and $c$ is 
{\color{black}the elastic wave velocity}
in the idealized plate.
Bear in mind that the first two parameters identify the sliding stability, although the last one turns out to be irrelevant.
In the following discussion, we restrict ourselves to the region where $\xi_2+v_l=v$ remains positive, and thus, $\mathrm{sgn}(\xi_2+v_l)=1$.
In the positive condition, we may circumvent the nonanalytic point $\xi_2+v_l=v=0$ that might cause a numerical instability (see appendix A).

\begin{figure}
	\begin{center}
		\includegraphics[width=6cm]{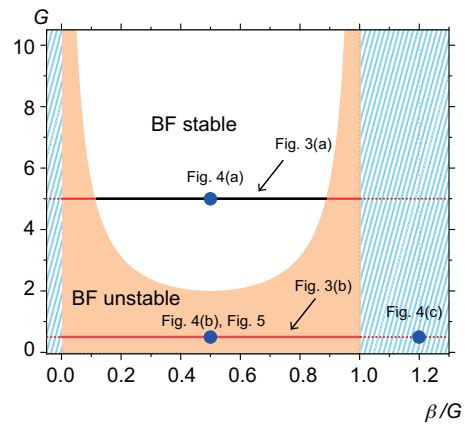}
		\caption{
		The line $1+(\mathrm{Im}\left[d \right]/ \mathrm{Re}\left[d \right])
		(\mathrm{Im}\left[g \right]/\mathrm{Re}\left[g \right])=0$ with respect to $G$ and $\beta/G$. The line indicates the boundary between the BF stable and the BF unstable regions denoted by orange shading.
		{\color{black}
		Regions with hatched lines in light blue are outside the application of Eq.~(\ref{BF_condition}).
        The thick horizontal lines or the blue dots are a visual guide to represent the boundary in the phase diagram of Fig.~\ref{G05-5} or data points in Fig.~\ref{G05-5_2}\,(a)-(c),\ and Fig.~\ref{slip_statics}, respectively.}
		}\label{amp}
	\end{center}
\end{figure} 

\section{complex Ginzburg-Landau equation}
\label{sec:wna}

\subsection{Hopf Bifurcation} We first determine a fixed point through a linear stability analysis without spatial coupling by setting $c^2 \partial^2 \xi_1/\partial x^2=0 $ in Eq.~(\ref{xi_time_evo}). The steady sliding is given by the solution $\bm{\xi}_s={}^t\!(\xi_{1s}, \xi_{2s}, \xi_{3s})$ to the stationary state $d\xi_i/dt=F_i(\bm{\xi})=0$:
\begin{eqnarray}
	\bm{\xi}_s
	&=&
	{}^t\left( -\frac{1}{G}\left(\frac{\mu_{*}}{a} - \beta\ln\frac{v_l}{v_*}  \right), 0, -(1+\beta)\ln\frac{v_l}{v_*} \right).
	\label{ls04}
	\end{eqnarray} 
We look at small deviations $\delta \bm{\xi}\equiv\bm{\xi}-\bm{\xi_s}$ around the steady-state solution to linearize
Eqs. \eqref{xi_time_evo}-\eqref{rs07} as 
\begin{eqnarray}\label{wna20}
\frac{d (\delta\xi_i)}{d t}=({\bf L})_{ij}\delta\xi_j,
\qquad \label{ls08-1}
	{\bf L}\equiv
	\left(
	 \begin{array}{ccc}
	 0 & 1 & 0\\
	 -G & -\frac{1}{v_l} & -1 \\
	 0 & -1-\beta & -v_l
	 \end{array}
	 \right).
\end{eqnarray}
The eigenvalues $\lambda$ of the matrix ${\bf L}$  satisfy the characteristic equation $ \det\left({\bf L}-\lambda {\bf I} \right)=0$.
Under the assumption that the roots of the cubic equation are one real root $\alpha$ and two complex roots $\epsilon \pm i \omega$,
the real root is obtained as $\alpha=-Gv_l/(\epsilon^2+\omega^2) <0$.
This ensures that the directions irrelevant to the Hopf bifurcation are stable (see appendix B).
In contrast, the bifurcation is signified by $\epsilon$.
At $v_l=v_c$, when $\epsilon$ turns from a negative to a positive value, a supercritical Hopf bifurcation appears with the following velocity and angular frequency:
\begin{equation}\label{ls12-2}
v_c=\sqrt{\frac{G-\beta}{\beta}}, \qquad \omega_c = \sqrt{G -\beta},
\end{equation}
which agrees with previous results~\cite{Rice1983}.
It is worth noting that only two crucial parameters $\beta$ and $G$ enter into the equation to determine the stability behavior.
The parameter $\beta$ is the distance from the velocity-weakening point, while $G$ represents the effective spring constant.
For Eq.~\eqref{ls12-2} to represent real values, $G-\beta>0$ and $\beta>0$, or $0<\beta<G$, is required. 
As long as this condition persists, the Hopf bifurcation point $v_c=\sqrt{(G-\beta)/\beta}$ always exists, and the non-oscillatory solution for $v_l>v_c$ is unstable. 

\subsection{CGLE}
We move on to a weakly nonlinear analysis of Eqs.~\eqref{xi_time_evo}-\eqref{rs07} to obtain the CGLE~(\ref{TDCGL}).  Bear in mind that the CGLE includes (i) a weakly nonlinear term and (ii) spatial local coupling.
As in Eq.~(\ref{wna20_1}), the governing equation is expanded in a power series with a finite order, which is referred to as ``weakly nonlinear analysis."
Loading velocity $v_l$ is set close to a Hopf bifurcation point $v_c$ as $v_l=v_c\left(1+\varepsilon \right)$
to expand the original Eq.~\eqref{xi_time_evo} in the Taylor series with the small deviation $\delta \bm{\xi}$ around the steady state:
\begin{eqnarray}\label{wna20_1}
\frac{d (\delta \bm{\xi})}{d t}=({\bf L} + {\bf D}\nabla^2)\delta \bm{\xi}+{\bf M}_0\delta \bm{\xi} \delta\bm{\xi} + {\bf N}_0\delta\bm{\xi} \delta\bm{\xi} \delta\bm{\xi}
+\cdots,
\end{eqnarray}
where
\begin{eqnarray}\label{wna21}
\left\{ 
\begin{array}{c}
{\bf L}_{ij}=({\bf L}_0)_{ij}+\varepsilon({\bf L}_1)_{ij}+\varepsilon^2({\bf L}_2)_{ij} \\
({\bf M}_0)_{ijk}=\frac{1}{2}\left. \frac{\partial^2 F_i}{\partial \xi_j \partial \xi_k}\right|_{{\bm \xi}={\bm \xi}_s} \\
({\bf N}_0)_{ijkl}=\frac{1}{6}\left. \frac{\partial^3 F_i}{\partial \xi_j \partial \xi_k \partial \xi_l}\right|_{{\bm \xi}={\bm \xi}_s}
\end{array}
\right. .
\end{eqnarray}
The matrix ${\bf L}$ is expanded in a series of $\varepsilon$ together with the $0$-th order ${\bf L}_0$ given by replacing $v_l$ with $v_c$ in the matrix of Eq.~\eqref{ls08-1}.
Here, we employ notation such as $({\bf N}_0\bm{u} \bm{v} \bm{w})_i=({\bf N}_0)_{ijkl} u_j v_k w_l$, with summations running over repeated indices.
In accordance with expansions in the matrices, the eigenvalue $\lambda$ is replaced with 
\begin{eqnarray}\label{wna22}
\lambda=\lambda_0+\varepsilon\lambda_1,
\qquad 
\label{wna23}
\left\{ 
\begin{array}{c}
\lambda_0
= i\omega_c\\
\lambda_1
=\frac{\beta  \sqrt{G-\beta }}{-\beta ^{3/2}+\left(\sqrt{\beta }-i\right) G}
\end{array}
\right.
\end{eqnarray}
A supercritical Hopf bifurcation is observed for ${\mathrm Re}[\lambda_1]>0$, which Eq.~\eqref{wna23} satisfies because we now just focus on $v_l>v_c \Leftrightarrow \epsilon >0$.
In addition, the diffusion (local coupling) term in the present system is specified by $({\bf D})_{ij}=c^2\delta_{i2}\delta_{j2}$.

Taking the term up to the $\varepsilon^{3/2}$ order with an appropriate procedure \cite{Kuramoto1984} (see also appendix G),
we arrive at the CGLE with the slow time variable 
$\tau = \varepsilon t$:
\begin{equation}
\frac{\partial W}{\partial \tau} =\lambda_1 W -g  |W|^2 W+d\frac{\partial^2 W}{\partial x^2}.\label{TDCGL}
\end{equation}
Note that the complex-valued $W(\tau,x)$ is associated  with the original governing Eq.~\eqref{xi_time_evo} through
\begin{equation}
\bm{\xi}(\tau,t,x)
=
\bm{\xi}_s
+ \sqrt{\varepsilon}
[\bm{\Xi} W(\tau,x) e^{i \omega_c t}+\mathrm{c.c.}],
\label{xi_W}
\end{equation}
where the right eigenvector for ${\bf L}_0$ with eigenvalue $\lambda_0$=$i \omega_c$ is given by
\begin{align}\label{wna28}
\bm{\Xi}
=
\left( 
\begin{array}{c}
-i\frac{1}{\sqrt{G-\beta}}  \\
1 \\
\frac{-\sqrt{\beta}+i \beta}{\sqrt{G-\beta}} 
\end{array}
\right)
\end{align} 
and the coefficients with complex values are

\begin{align}
g&=&\frac{-i \beta ^2 (\beta +1) (9 G-8\beta)}{6 \sqrt{G-\beta } \left[2 \beta  ( G- \beta)^2-3
	i \sqrt{\beta } G (G-\beta)-G^2\right]}
	\nonumber \\
	\label{wna56}
\end{align}
\begin{eqnarray}\label{diffusion_d}
d 
=
\frac{c^2 \left(i\sqrt{\beta }+1\right) \sqrt{G-\beta }}{2 \left[\beta ^{3/2}-\left(\sqrt{\beta
	}-i\right) G\right]}.
\end{eqnarray}
The time scale between $t$ and $\tau$ is separated enough for the slow variable $W(\tau,x)$ to undergo an adiabatic change. 
Equation~\eqref{TDCGL} is a general form for a CGLE obeying the slow variable $\tau$. 
The universal spatio-temporal structures are captured by the squared or cubed term of $\delta \bm{\xi}$
in the weakly nonlinear regime.
For example, the general form of Eq.~\eqref{TDCGL} includes the Ginzburg-Landau equation with the real coefficients $Im[\lambda_1]=Im[g]=Im[d]=0$.
This is very often exploited to phenomenologically discuss phase transition dynamics with a variational function (Landau free energy), during which a system develops towards a minimum point.
In contrast, if $\lambda_1=Re[g]=Re[d]=0$, we encounter the nonlinear Schr$\ddot{{\rm o}}$dinger equation
with Eq.~\eqref{TDCGL}~\cite{Aranson2002}, where conservative nonlinear wave phenomena are found  without a dissipation mechanism.
The present system is between these regimes and
identified with a combination of the complex coefficients $g$ and $d$.

Before including spatial coupling, let us discuss the general role of the complex coefficient $g$ of the cubic term through a homogeneous example with $d=0$, or equivalently, the one reduced to the Stuart-Landau equation.
Qualitative features
are grasped in the amplitude-phase representation by 
 rephrasing Eq.~\eqref{TDCGL} with $W=R e^{i\Theta}$:
\begin{align} 
\frac{d R}{d\tau}=Re [\lambda_1]R-Re [g]R^3, \label{Req} \\
\frac{d \Theta}{d\tau}=Im[\lambda_1]-Im [g]R^2,\label{Thetaeq} \end{align}
where amplitude $R$ and phase $\Theta$ are real numbers.
Eq.~\eqref{Req} gives the steady solution $R_s=\sqrt{Re[\lambda_1]/Re[g]}$, which implies that the solution to the original Eq.~\eqref{xi_time_evo} oscillates with a finite amplitude. 
The cubic term in Eq.~\eqref{TDCGL} provides the solution not obtained from the linear analysis with Eq.~\eqref{wna20}.
In addition, a close inspection shows that a leading part of the angular frequency is determined by $\omega_c$, but modified by $d \Theta/d \tau$ [Eq.~\eqref{Thetaeq}]. Notably, the oscillation amplitude $R$ affects the angular frequency through a higher-order perturbation term. 

\subsection{Benjamin-Feir Instability}
\begin{figure}
	\begin{center}
		\includegraphics[width=6.5cm]{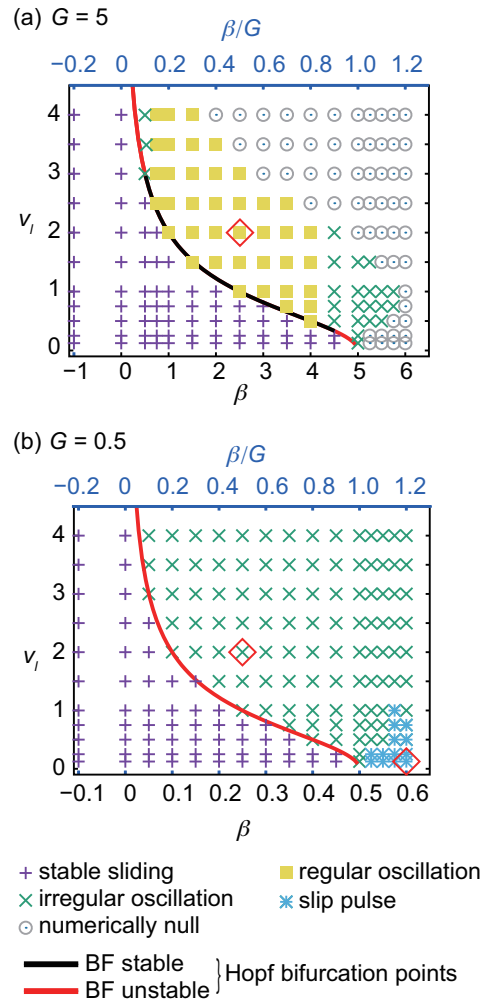}
		\caption{Phase diagram {\color{black}on the $\beta$-$v_l$ plane} built from Eqs.~\eqref{eq-rsall}-\eqref{rs07} for (a) $G$=5 and (b) $G$=0.5. Solid lines represent the analytical results for Hopf bifurcation points $v_c$ with Eq.\eqref{ls12-2}.
		Black and red curves indicate the absence and presence, respectively, of the BF instability based on Eq.~\eqref{BF_G_beta}. 
		{\color{black}Upper horizontal axes indicate $\beta/G$ reduced from $G$ and $\beta$, and}
		red diamonds indicate the points for the example given in Fig.~\ref{G05-5_2}\,(a-c).
		}
		\label{G05-5}
	\end{center}
\end{figure}

When (ii) spatial local coupling participates in Eq.~\eqref{TDCGL}, spatio-temporal chaos, referred to as the BF instability, emerges for 
\begin{eqnarray}\label{}
&& 1+\frac{Im\left[d \right]}{Re\left[d \right]} \cdot \frac{Im\left[g \right]}{Re\left[g \right]}
\nonumber \\
&=&
\frac{[2 \beta ( G-\beta) - G] [\beta (G-\beta)^2+G^2] }{3 \beta^2 G (G-\beta) }<0,
\label{BF_condition}
\end{eqnarray}
where the spatial gradient of the phase distribution is enhanced.
The BF instability appears when the gradient in the phase modifies the amplitude $R$, and it can be intensified through Eq.~\eqref{Thetaeq}. 
Indeed, in the original Eq.~\eqref{xi_time_evo}, we find that the coupling term $c^2 \partial^2 \xi_1/\partial x^2$ appears in the time derivative of $\xi_2$, not in $\xi_1$.  The coupling term does not smooth the phase but shrinks the amplitude when the phase advances more than those of neighboring locations;  consequently, the BF instability may appear.
From the positive nature of  $\beta, G$, and $G-\beta$, 
the condition for the BF instability is reduced to 
\begin{equation}
G < \frac{1}{2\frac{\beta}{G}(1-\frac{\beta}{G})},
\label{BF_G_beta}
\end{equation}
which separates the BF-stable and BF-unstable regions in the {\color{black}$G$-$\beta/G$ plane, as shown in Fig.~\ref{amp}. 
{\color{black}
The boundary curve exhibits a convex downward shape and diverges at $G/\beta=0,\,1$.
Along the lower thick line as a function of $\beta/G$ at a fixed small $G=0.5$ below the minimum point of the boundary curve, the system is always unstable from $G/\beta=0$ to $1$ in the absence of a sufficient restoring force.
This is a remarkable result because it implies that chaotic behavior appears if the material is soft enough.}
In contrast, if $G$ is large enough, e.g., $G=5$, the system may enter the BF-stable regime at an intermediate value of $\beta/G$.}
We also use Eq.~\eqref{BF_G_beta} to distinguish the BF stability from BF instability by coloring the boundary curves with red and black, respectively, on the phase diagram in Fig.~\ref{G05-5}.

\begin{figure*}
	\begin{center}
		\includegraphics[width=17.2cm]{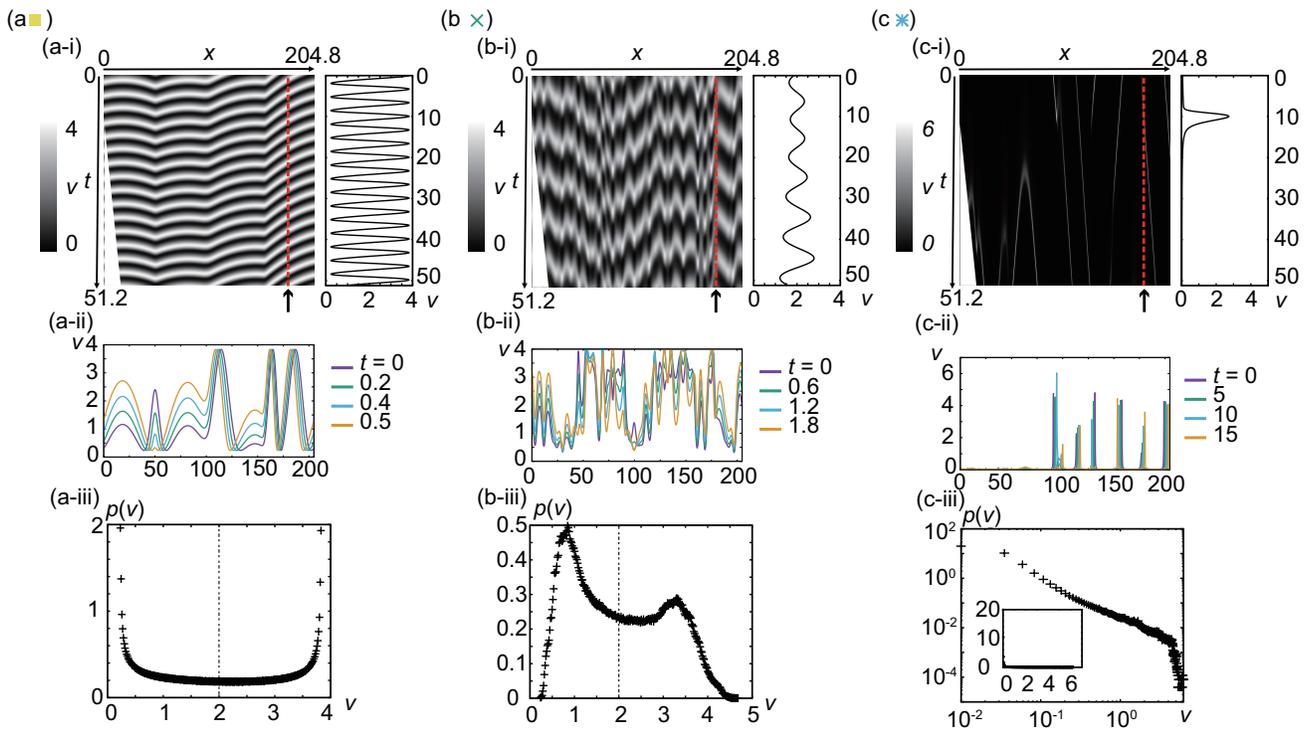}
		\caption{Numerical results obtained from Eqs.~\eqref{eq-rsall}-\eqref{rs07}: (a) regular oscillation, (b) irregular oscillation, and (c) slip pulse. 
		(i) Spatio-temporal plots.
		Slopes of white triangular areas at the lower left corners on the $t$-$x$ planes represent {\color{black}
		the elastic wave velocity.
		}
		Temporal history of $v$ along $t$ (red dashed lines on the right side of each plot).
		(ii) Spatial plots of the slip velocity. 
		(iii) Probability density functions of observed slip velocity $p(v)$. Dotted lines in (a) and (b) indicate $v_l$.  Parameters are (a) $G=5$, $\beta=2.5$, $v_l=2$ {\color{black}($\beta/G=0.5$)}; (b) $G=0.5$, $\beta=0.25$, $v_l=2$ {\color{black}($\beta/G=0.5$)}; and (c) $G=0.5$, $\beta=0.6$, $v_l=0.125$ {\color{black}($\beta/G=1.2$)}. 
		}
		\label{G05-5_2}
	\end{center}
\end{figure*}

\section{Numerical results}	
\label{Numerical}
We performed a numerical simulation using the original governing Eqs.~\eqref{eq-rsall}-\eqref{rs07} to verify the reductive perturbation predictions.
Note that an auxiliary viscous friction linearly proportional to the relative velocity between adjacent segments $\nu \partial^2 \xi_2/\partial x^2$ was incorporated as the second component of Eq.~\eqref{eq-rsall} to suppress the numerical instability. 
Indeed, appendix G provides positive evidence that the appropriate viscous friction was not strong enough to alter the qualitative behavior.

An explicit scheme under a periodic boundary condition was employed.
Bear in mind that, as mentioned earlier, the simulation was halted when the velocity took on an invalid value $v=\xi_2 + v_l <0$ [gray circled dots in Fig.~\ref{G05-5}\,(a)]. 
The simulation started at $t=0$, when the initial condition was prepared from the homogeneous steady solution of Eqs.~\eqref{rs05}-\eqref{rs07} with the addition of small spatio-temporal disturbances.
Because a strong artifact from the initial random conditions is unfavorable, we waited for the system to settle, and then the numerical samples were collected (see appendix C for detailed conditions).

\subsection{Phase Diagram}

Typical phase diagrams are given in Fig.~\ref{G05-5}\, (a),\,(b), where 
Hopf-bifurcation points sweep along solid curves determined with Eq.~\eqref{ls12-2}. 
In fact, persistent oscillation is not seen below these curves. 
The black and red curves signify whether the BF instability is absent or present, respectively.

Yellow squares around the black curve in Fig.~\ref{G05-5}\,(a)
indicate regular oscillations, whose dynamic behavior is exemplified in Fig.~\ref{G05-5_2}\,(a-i). 
{\color{black}The pattern that appeared was not propagation but rather 
almost uniform oscillation with gradual spatial variation of the oscillation phase.}
Incidentally, the translational symmetry of the pattern was broken as a consequence of the initial randomness.
Additionally, Fig.~\ref{G05-5_2}\,(a-i) shows the velocity cross section taken from the density diagram along the red broken dashed line.
The velocity profile looks symmetric about the loading velocity ($v_l=2$).
More quantitative analyses are shown in Fig.~\ref{G05-5_2}\,(a-iii)\,(bottom).
The duration of time spent around the maxima or minima of $v$  
was the longest, which means that $p(v)$ has peaks near $v=0$ or $v=4$.

In contrast, irregular oscillations were observed (green crosses) near the red curves in Fig.~\ref{G05-5_2}\,(b).
The typical behavior of irregular oscillation has instability at short wavelengths, as in Fig.~\ref{G05-5_2}\,(b-i),\,(b-ii).
The wave pattern propagated at the velocity indicated by the white triangular area, which corresponds to the sound speed. This irregular oscillation meets with an example of spatio-temporal chaos due to the BF instability, because the phase distributions have random spikes in Fig.~\ref{G05-5}\,(b-ii).
A velocity profile was also taken from the spatio-temporal density map along the red dashed line.
Although the oscillations are centered at $v=2$, amplitudes became smaller than those of Fig.~\ref{G05-5}\,(i). 
This point is made clear by looking at $p(v)$ in Fig.~\ref{G05-5}\,(iii), where bimodal distributions are found.
In addition, the duration time spent around the lowest velocity was longest for this condition. Thus, a peak around $v\simeq 1$ is found, and $p(v)$ turns out to be asymmetric.

The other phase is a slip pulse discovered at the points marked by blue asterisks in Fig.~\ref{G05-5}\,(b). 
In the  parameter region, spatially localized domains with finite slip velocity propagated at the sound speed, as those have a slope comparable to that of the white triangular area at the lower left corner. 
The slip velocity distributions look like a power law, where a large number of events on the distribution was found at small $v$. 
The qualitative condition for pulse occurrence is similar to that in previous studies, where the slip pulse was reported for the model with velocity weakening friction~\cite{Hirano2016, Ampuero2008}.

\begin{figure*}
	\begin{center}
		\includegraphics[width=17cm]{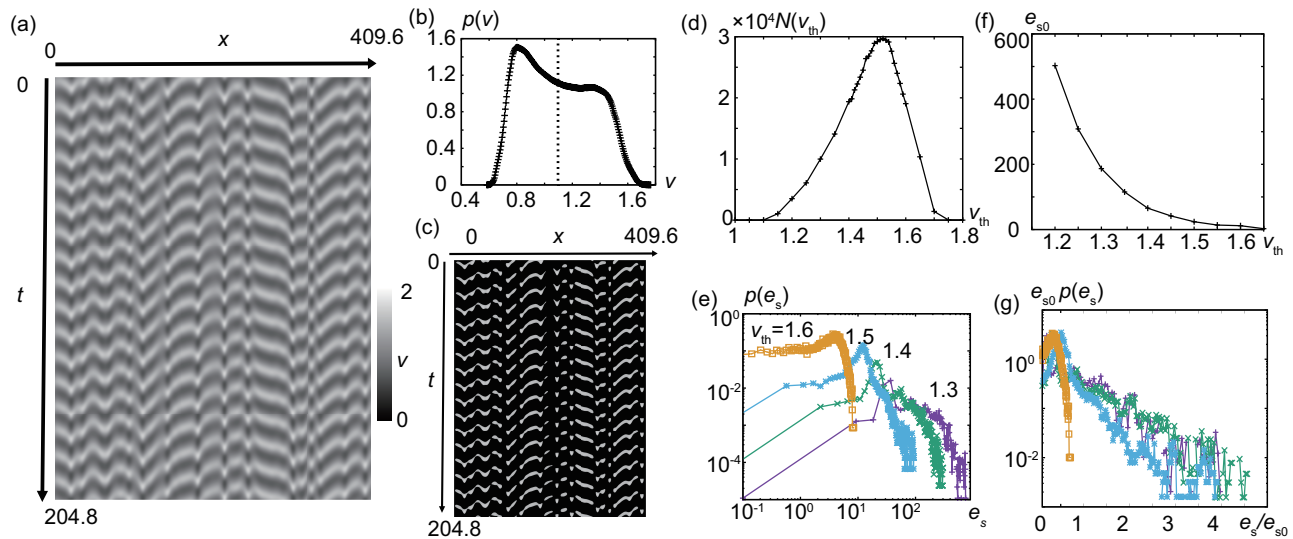}
		\caption{Numerical results relevant to slip event size with $G=0.5$, $\beta=0.25$, and $v_l=1.1$ {\color{black}($\beta/G=0.5$)}. (a) Spatio-temporal plot of irregular oscillations.
		(b) Distribution of the slip velocity $p(v)$, with $v_l=1.1$ represented by the dashed line. 
		(c)
		Binarized plot of (a) separated by $v_{th}=1.4$. The gray scale bar for $v$ is the same as that in (a), except that values for $v<v_{th}$ are replaced with solid black. 
		(d) Number of isolated slip events $N(v_{th})$ depending on $v_{th}$. (e) Probability density function of slip event size $p(e_s)$ for various threshold velocities $v_{th}=$1.3 (purple), 1.4 (green), 1.5 (cyan), and 1.6 (brown). 
		(f) Typical event sizes $e_{s0}$ estimated from $p(e_s)\sim \exp{(-e_s/e_{s0})}$ are plotted against $v_{th}$.
		(g) Rescaled plots of $e_s p(e_s)$ with $e_s/e_{s0}$. The same symbols as in (e) are adopted. 
		}
	\label{slip_statics}
	\end{center}
\end{figure*}

Focusing attention on the irregular oscillation associated with the BF instability among these four types of characteristic dynamics, 
we look for common features shared with slow earthquakes, such as tectonic tremors or slow slip events. 
One of the noteworthy points is the event size distribution. 
The simulation was performed in a larger space and with longer duration to collect statistical data [see Fig.~\ref{slip_statics}\,(a) and appendix C for detailed conditions].
The slip behavior is exemplified in Fig.~\ref{slip_statics}\,(a), where 
an irregular oscillation is observed.
Figure~\ref{slip_statics}\,(b) shows the asymmetric bimodal distribution of slip velocity $p(v)$ with respect to the loading velocity $v_l$.

\subsection{Event size distributions}
\label{chap_event_size}

{\color{black}
One of the important indices is the cumulative distribution of separate slip events.
To identify separate events on a continuous spatio-temporal map,
we binarize the density plot separated by threshold velocity $v_{th}$ that distinguishes ``slipping'' for $v>v_{th}$ from ``no slipping'' for $v<v_{th}$.}
The connected slipping domains on a spatio-temporal {\color{black}map,} that is, the area enclosed by a contour line on a spatio-temporal diagram [see the white-gray domain in Fig.~\ref{slip_statics}\,(c) with $v_{th}=1.4$, below which the regions are drawn in solid black], are considered as a single slip event size $e_s$.

The number of observed slip events $N(v_{th})$ varies according to $v_{th}$ [Fig.~\ref{slip_statics}\,(d)]. 
For $v_{th}<1.1$, all the areas are completely connected and counted as a single slip event, which does not make sense statistically. However, for $v_{th} > 1.7$, no slip event is detected over the whole spatio-temporal plane.
In the range $1.1<v_{th} < 1.7$, we monitored the distribution of event size for various $v_{th}$ values. 
Figure~\ref{slip_statics}\,(d) has a single peak around $v_{th}=1.5$.

The shapes of isolated slip event domains do not look like fractal structures.
This fact is verified by looking at the consistency with the Mandelbrot conjecture, which states that, if Korczak's empirical law $N(v_{th}) \sim (v_{th})^{\zeta_N}$ holds, 
 the number of the isolated domains $N(v_{th})$ is supposed to be associated with the contour fractal curves characterized by the Hurst exponent $H$ through  $\zeta_N=1-H/2$~\cite{Mandelbrot1982, Mandelbrot1968}.
In particular, this was discussed for the fractal contours drawn with fractional Brownian motion~\cite{Matsushita1991}.
This is not, however, true in the present system.
As a matter of fact,
neither the appearance of  Fig.~\ref{slip_statics}\,(c) nor the distribution in Fig.~\ref{slip_statics}\,(d) seem to provide exponents $H$ and $\zeta_N$. 

The probability density function $p(e_s)$, with $e_s$ estimated from binarized plots like those in Fig.~\ref{slip_statics}\,(c),
is shown in Fig.~\ref{slip_statics}\,(e).
As $v_{th}$ changes, exponential decreases along the event size $e_s$ are maintained.
We are also aware that it is unlike the power law. 
Such exponential distributions 
resemble the cumulative distributions of seismic energy rates observed in {\color{black}tectonic tremors associated with slow earthquakes~\cite{Yabe2014}.} The exponential distributions are meant to have their own characteristic sizes. 
Fitting the exponential decay $p(e_s)\sim \exp{(-e_s/e_{s0})}$ on Fig.~\ref{slip_statics}(e),  we estimated 
the characteristic
event sizes $e_{s0}$, which are plotted in Fig.~\ref{slip_statics}~(f).
In addition, Fig.~\ref{slip_statics}(e) is rescaled with $e_{s0}$, as in Fig.~\ref{slip_statics}~(g).

\section{Discussion}
\label{Discussion}

Numerical and analytical aspects are discussed in secs.~\ref{sec:wna} and \,\ref{Numerical}, respectively.
Let us first verify the agreement between them from the viewpoint of qualitative and quantitative consistencies. 

The qualitative point is the distribution shape.
Recall that the numerical event size displays the exponential distribution, where the temporal plots on the spatio-temporal plane are fairly periodic [see the right side profile of Fig.~\ref{G05-5_2}\,(b-i)], but the spatial plots are disordered [see Fig.~\ref{G05-5_2}\,(b-ii)].
To make the point clear, we address both the temporal and spatial aspects.
Although the temporal period varies over the long term with spatio-temporal chaos, a periodic pattern is found, implying that the characteristic time may correspond to the temporal period $T_c$.
In contrast, the spatial profiles are disordered rather than periodic because the neighboring differences in phase are intensified due to the BF instability mechanism.
Especially, the spatial correlations get lost in the finite distance $\Lambda_c$, suggesting that the correlations decay in an exponential manner. 
Therefore, the characteristic time and length may be defined using the temporal period and correlation length, respectively.
Because the event size is defined as its area (time $\times$ length) on the spatio-temporal plane, it is a natural consequence if an event has the characteristic event size $e_{sc}\sim T_c \Lambda_c$, for which the size distribution is exponential.

Let us next review the quantitative applicability.
As shown in sec.~\ref{sec:wna}, the characteristic time should be comparable to 
$T_c=2 \pi/\omega_c=2 \pi/\sqrt{G-\beta}$. 
The characteristic length $\Lambda_{c}$ is estimated from the wavelength of the most unstable mode on the propagating wave solution (see appendix D): 
\begin{equation}
\Lambda_{\mathrm{c}} 
= 
2 \pi \sqrt{\frac{Re [d]}{Re[\lambda_1]}}
\sqrt{-\frac{(\frac{Im [d] }{Re[d] })^2 (1+ (\frac{Im [g] }{Re[g ] })^2) }{1+\frac{Im\left[d \right]}{Re\left[d \right]} \cdot \frac{Im\left[g \right]}{Re\left[g \right]}}  }.
\label{Lambda_c}
\end{equation}
{\color{black} These characteristics indicate} that event sizes are estimated with only two parameters, $G$ and $\beta$.
Setting the parameter values as $G=1/2$ and $\beta=1/4$,
{\color{black}employed} in Fig.~\ref{slip_statics},
we estimated the event size $T_c \Lambda_{c} \simeq 100$ with 
$T_c=4\pi$ and $\Lambda_{c}=3\pi$.
The analytically estimated event size is on the same order of the size obtained from the numerical results in Fig.~\ref{slip_statics}\,(f). Thus, the analytical results with the CGLE show qualitatively and quantitatively excellent agreement with those of the numerical simulation.

{\color{black}A verification by laboratory experiments is necessary to determine
the CGLE validity.
Guided by Fig.~\ref{G05-5} and Eq.~\eqref{BF_G_beta}, the slip mode becomes chaotic for
small values of $G$, which indicates that soft solids such as rubber and gel are promising materials for laboratory experiments.
In a soft material sheet made from gels~\cite{Baumberger2002, Baumberger2006, Yamaguchi2009a,Baumberger2003, Fukudome_},
}
{\color{black} the estimated values for $\hat{\Lambda_c} \simeq $ $10^{-1}$~m and $\hat{T}_c \simeq$ $10^{-2}$~s {\color{black} would cause irregular slip due to the BF instability (see Table I in appendix E)..}
Although the characteristic length $\Lambda_c$ is slightly large, the experimental conditions fall into the feasible length and timescale of the observation for the BF instability by adding some modifications, such as different values for the rigidity, thickness, and loading speed. {\color{black}In addition to gels, rubber sheets ($\hat{G}\simeq 10^{6}$ Pa) with a thickness on the order of millimeters ($\hat{h}\simeq 10^{-3}$~m) are also promising candidates because they have $\hat{\Lambda_c} \simeq $~$10^{-2}$~m and $\hat{T}_c \simeq$ $10^{-3}$~s.}

In addition, let us here remark on the relation between the elasticity and the thickness. 
We are aware that the thickness is one of the important parameters because the elastic interaction has a range comparable to that of the thickness.}
{\color{black}Probably two limiting cases have been investigated most frequently:
a thin layer or semi-infinite elastic half-space.
However, we are not sure if the elastic-interaction range can alter the qualitative observations. In soft solids like rubber and gels, the thickness is rather easy to adjust so as to evaluate both thin and thick plates, whereas the present mathematical model constructed in the thin layer deals only with local coupling.
The pertinent phenomena do not seem to have been observed yet at the laboratory scale, but the conditions described above would offer feasible projects.}

{\color{black} Other practical applications are brake pads~\cite{Behrendt2011}, windscreen wiper blades~\cite{Lancioni2016}, and tires~\cite{Persson1997}, where a relatively high loading rate as well as a thin elastic body are present. In such cases, spatially synchronized oscillation is undesirable in terms of stability, and the introduction of the spatio-temporal chaos due to BF instability may prevent such coherent oscillation.
We may also include the peeling dynamics of soft adhesives~\cite{Dalbe2015, Dalbe2016}, where micro-scale stick-slip motion occurs, as a possible candidate for our analysis.}

We then move on to a discussion about slow earthquakes.
Recent seismic observations~\cite{Yabe2014} have shown that the cumulative distribution of the seismic energy rate can be fitted well using an exponential distribution.
As seen in Section~\ref{chap_event_size}, the CGLE around the BF instability closely reproduces this exponential feature, which is in ``qualitative" agreement with the observed consequences~\cite{Yabe2014}.
However, a ``quantitative" agreement has not been achieved.
Indeed, when plausible parameters~\cite{kano2010, Liu2005, Scholz1998} are plugged into the CGLE, we encounter unrealistically huge scales for the characteristic length $\hat{\Lambda}_{\mathrm{c}} \simeq 10^{19}~\mathrm{m}$ and time $\hat{T}_c \simeq$ $10^{7}$~s (see Table I in appendix E).
Nonetheless, we should not rush to the conclusion that the CGLE approach is not completely appropriate in the study of slow earthquakes, because the CGLE approach itself is a very general framework independent of the specific structure of the system. 
Tracing back the derivation with the reductive perturbation method, the quantitative difficulty begins at Eq.~(\ref{ls12-2}), where {\color{black}the critical velocity $v_c=v_l$} is too low to meet with an appropriate oscillation period, as seen from $\hat{T}_c=2\pi/\hat{\omega}_c=2\pi/(v_l\sqrt{\beta})$. Thus, {\color{black}
the primary modification should lie in a starting point around Eqs.~\eqref{xi_time_evo}-\eqref{rs07}
rather than the coupling term because Eq.~(\ref{ls12-2}) is a result obtained as an independent oscillator.
This means that improvement from the local to nonlocal coupling~\cite{Tanaka2003} is not enough to reproduce the realistic order estimate, although the thin layer is certainly a useful approximation.
Alternatively, we arrive at the other possible candidates to improve the situation.
Faster oscillation may be triggered by additional {\color{black}hidden} variables, different from the elastic origin. 
For instance, recent observations reveal that fluid migration and precipitation of dissolved chemicals may control slow earthquakes~\cite{Audet2014}.
They could bridge the quantitative gap while retaining the quantitative manner because the spatio-temporal chaos due to the BF instability derived from the general framework of the CGLE maintains an exponential dependence.

In addition to slow earthquakes, we suggest possible applications of our model to other geologically meaningful regimes. Such situations can be realized with increased loading velocity. 
Indeed, $\hat{v}_l \simeq$ 1~m/s around the onset velocity of ordinary earthquakes that start to release seismic waves provides $\hat{\Lambda}_c \simeq 10^3$ m or even a smaller order of magnitude for smaller $\hat{h}$. This implies that chaotic slip may occur inside faults on a subducting plate.
}

{\color{black}
\section{Concluding Remarks and Perspectives}
\label{Conclusion}

We have discussed the application of the CGLE to the oscillatory instability observed in the thin layer model with the RSF law.
The CGLE has a long history of being employed as a successful framework near the bifurcation (transition) points for various phenomena.
Thus, this approach could embrace a diverse range of unstable interface systems.

Our analytical and numerical studies have primarily investigated the BF instability leading to chaotic behavior in light of the CGLE, and then
we applied it to two notable cases in the main text: soft matter and slow earthquakes.
To our knowledge, pertinent laboratory experiments with soft solids have not been reported.
Instead, we have proposed feasible conditions for experiments with rubber or gel.
{\color{black}A close inspection of the $G$-$\beta/G$ diagram in Fig.~\ref{amp} implies that soft solids are promising for verifying chaotic behavior in laboratory experiments due to their small compliance.
One of the authors revealed the subsonic to intersonic transition in sliding friction of silicone gels~\cite{Yashiki2020}. According to the proposal in our study, friction of soft solids can exhibit spatio-temporal chaos with a loading velocity on the order of 10$^{-2}$ m/s or higher. Such chaotic behavior may prevent tires from entering undesirable oscillatory synchronization for the practical purpose of braking. We emphasize that high-speed friction of soft matter should contain rich and fruitful physical phenomena to be investigated.

Slow earthquakes are also considered as an applicable issue.}
Comparing the observations of slow earthquakes with the analytical and numerical results, we can discover the qualitative coincidence of the exponential dependence of event size,
{\color{black}whereas the quantitative estimates provide different results.
The discrepancy arises from the fact that the present model lacks potentially crucial elements, such as heterogeneity or pore fluid pressure.
Modifications by incorporating these elements could improve the quantitative gap without changing the qualitative aspects. This speculation is reasonable because the exponential distribution of slip event size is predicted by the BF instability mechanism derived from the general framework of the CGLE.}
Considering the broad ranging applicability of the CGLE, further studies based on the CGLE are expected to 
contribute to the understanding of slow earthquakes, as well as the slip instability occurring in soft matter.
}

\subsection*{Acknowledgment}
	This work was supported by the Japan Society for the Promotion of Science (JSPS), KAKENHI Grants JP16K13866, JP16H06478, {\color{black}JP19H05403, and JP21H05201}.
	This work was also supported by JSPS and PAN under the Japan-Poland Research Cooperative Program ``Spatio-temporal patterns of elements driven by self-generated, geometrically constrained flows,'' and the cooperative research of ``Network Joint Research Center for Materials and Devices'' with Hokkaido University (Nos.~20161033, 20171033, and 20181048).

\section*{Appendix}

\subsection*{A. Constraints on the sign of $v$}
It should be noted that, when $\xi_2+v_l=v=0$, Eqs.~\eqref{rs06} and \eqref{rs07} are not analytic.  We tried several modifications of Eqs.~\eqref{rs06} and \eqref{rs07} to avoid the non-analytic behavior near $v=0$, especially to stabilize  the numerical calculation. However, violent oscillation appeared,
even after modifications,
when $v$ changed its sign. Resolving this situation is beyond the scope of this study. Thus, we analytically and numerically restricted ourselves to 
$\xi_2+v_l=v>0$. 
If $\xi_2+v_l=v$ became 0 or negative,
the numerical simulation was halted.

\subsection*{B. Linear analysis}
When a cubic equation has one real root, $\alpha$, and two complex roots, $\epsilon \pm i \omega$, it is written as
\begin{equation}\label{ls10}
\lambda^3-\left(\alpha + 2 \epsilon \right)\lambda^2+\left(2 \alpha \epsilon + \epsilon^2 + \omega^2 \right)\lambda -\alpha \left(\epsilon^2+\omega^2 \right)=0.
\end{equation}
Comparing the characteristic equation $ \det\left({\bf L}-\lambda {\bf I} \right)=0$ reduced to
$\lambda^3+\left(v_l+1/v_l  \right)\lambda^2+\left(G-\beta \right)\lambda +G v_l=0$ with Eq. (\ref{ls10}), we obtained
\begin{eqnarray}
\alpha + 2\epsilon&=&-v_l-1/v_l 
\label{ls11}
\\
\epsilon^2+\omega^2+2\alpha \epsilon &=& G -\beta
\label{ls11-2}
\\
-\alpha \left( \epsilon^2+\omega^2 \right)&=&Gv_l.
\label{ls11-3}
\end{eqnarray}
The parameter region $\beta>0$ and $\beta >G$ finds the critical velocity and angular frequency, respectively, in Eq.~\eqref{ls12-2}, as in the main text.

In contrast, if $\beta<0$, oscillation does not appear in any sliding velocity $v_l$.
Also, if $\beta>G$, oscillation emerges for any positive sliding velocity $v_l$. 
The reductive perturbation method cannot be applied near the bifurcation point in either case {\color{black}(see also Fig.~\ref{amp})}, but they are not of interest here.

\subsection*{C. Numerical setup}
An explicit scheme under a periodic boundary condition was employed {\color{black}in the numerical simulations}.
The system was discretized with sizes $\Delta t=0.0002$ and $\Delta x=0.05$. In addition, the dimensionless parameters $c^2 = 0.1$, $\nu =0.01$, $\sigma^{*}=1$, and $v^{*}=0.1$ were used in {\color{black}Figs.\ref{G05-5} and \,\ref{G05-5_2}.}
The simulation was halted when the velocity assumed an invalid value, or $v=\xi_2 + v_l >0$. 
Although a time evolution that obeys the CGLE is deterministic,  
random initial conditions were generated by adding small disturbances to the homogeneous steady solution of
{\color{black}Eqs.~\eqref{xi_time_evo}-\eqref{rs07}.}
In Figs.~\ref{G05-5} and~\ref{slip_statics}, the system sizes were 204.8 and 409.6, respectively. 

Each simulation started at $t=0$ from a random initial condition.
Because a strong artifact due to the initial condition is unfavorable, we waited for the system to settle. 
In Fig.~\ref{G05-5}, data were collected after $t=4000$. The sampling was done with $\delta x =0.2$ and $\delta t = 0.05$ to obtain 1024 $\times$ 1024 spatio-temporal data points. 
{\color{black}Similarly,}
for the simulation shown in Fig.~\ref{slip_statics}, data were collected after $t=8427.6$. The sampling was done with $\delta x =0.2$ and $\delta t = 0.1$ to obtain 2048 $\times$ 163,920 spatio-temporal data points. 

The numerical codes can be obtained from~\cite{dummy2}.

\subsection*{D. Instability wavelength}
Numerical calculations performed with the original governing equation produced irregular oscillation with the parameters used, whereas the BF instability was predicted by the analytical study. Furthermore, it was revealed that the event size defined as the area with a threshold slip velocity $v_{th}$ exhibits an exponential distribution. This exponential distribution suggests that there is a typical event size $e_{sc}$, and $e_{sc}=T_c \Lambda_c$ is written with a typical time $T_c$ and length $\Lambda_c$. As discussed in sec.~\ref{sec:wna}, the oscillation period is comparable to $T_c$, and the analytical estimate leads to $T_c=2 \pi/\omega_c=2 \pi/\sqrt{G-\beta}$.
In contrast, the typical spatial size $\Lambda_c$ is estimated by the correlation length on the BF instability.
To give a specific analytical expression for $\Lambda_c$, we here turn back to the CGLE~(\ref{TDCGL}):
The solution of the equation describes a propagating wave: 
\begin{equation}
W_k = \sqrt{\frac{Re [\lambda_1]}{Re [g]}} \sqrt{1-\frac{Re [d]}{Re [\lambda_1]}k^2}\exp(ikx+\omega_k t)\label{TDGLsol}
\end{equation}
and
\begin{equation}
\omega_k = Im[\lambda_1] -\frac{Im[g]}{Re[g]}Re[\lambda_1]+\left(\frac{Im[g]}{Re[g]}Re[d]-Im[d] \right)k^2 \label{TDGLsol2}.
\end{equation}
The solution exists for $k<k_{\mathrm{max}}$ or $\lambda > \lambda_{\mathrm{min}}$ with $k_{\mathrm{max}}=2\pi/\lambda_{\mathrm{min}}$,
where
\begin{equation}
\lambda_{\mathrm{min}} =2 \pi \sqrt{\frac{Re [d]}{Re[\lambda_1]}}
=2 \pi \sqrt{\frac{c^2}{2(G-\beta)}}
\label{k_max}.
\end{equation}
Moreover, for the propagating wave to be stable  against infinitesimal perturbations, we require an additional condition:
\begin{eqnarray}
\lambda
&>&
\lambda_{\mathrm{min}} \sqrt{\frac
{3 +\frac{Im\left[d \right]}{Re\left[d \right]} \cdot \frac{Im\left[g \right]}{Re\left[g \right]}  +2 (\frac{Im [g] }{Re[g ] })^2}
{1+\frac{Im\left[d \right]}{Re\left[d \right]} \cdot \frac{Im\left[g \right]}{Re\left[g \right]}}  }.
\end{eqnarray}
{\color{black}When $1+(Im [d]/Re [d])(Im [g]/Re[g]) <0$, even homogeneous oscillation $k=0$ becomes unstable against infinitesimal perturbation,  namely the BF instability. Specifically, the most unstable wavelength [Eq.~(\ref{Lambda_c})] is written as
\begin{eqnarray}
\Lambda_c
&=&
\lambda_{\mathrm{min}} \sqrt{\frac{[G+\beta(G-\beta)]^2 [G^2+4\beta (G-\beta)^2]}{3\beta^2 G (G-\beta) [G-2\beta(G-\beta)]}}
\label{lambda_max_stable}.
\end{eqnarray}}
{\color{black}Note that the typical spatial size $\Lambda_c$ is estimated from the most unstable wavelength.}
Eventually, we arrive at the analytical estimate for $e_{sc}=T_c\Lambda_c$ with $T_c=2\pi/\omega_c$, Eq.~(\ref{ls12-2}), and Eq.~(\ref{lambda_max_stable}).

When the value used for the numerical simulation in Fig.~\ref{slip_statics} is inserted, $T_c = 4 \pi$, 
$\Lambda_{c}=3\pi$, and $e_{s0} = 12 \pi^2 \simeq 100$.

\begin{table}[t]
\centering
\begin{tabular}{|c|c|c|c|}
\hline
& gel & rubber & slow EQ \\ \hline \hline$\hat{\rho}$ & $10^{3}$~kg/m$^3$&$10^{3}$~kg/m$^3$ & $10^{3}$~kg/m$^3$ \\ \hline
$\hat{G}$ & $10^{5}$~N/m$^2$&$10^{6}$~N/m$^2$&$10^{10}$~N/m$^2$  \\ \hline
$\hat{\sigma}$ & $10^{3}$~N/m$^2$ & $10^{3}$~N/m$^2$ & $10^{8}$~N/m$^2$  \\ \hline
$a\hat{\sigma}$ & $10^{2}$~N/m$^2$ & $10^{2}$~N/m$^2$ & $10^{6}$ N/m$^2$   \\ \hline
$\hat{v}_l$ & $10^{-2}$~m/s & $10^{-2}$~m/s & $10^{-9}$~m/s  \\ \hline
$\hat{D}_c$ & $10^{-2}$~m & $10^{-3}$~m & $10^{-2}$~m   \\ \hline
$\hat{t}_u$ & $10^{-2}$~s & $10^{-3}$~s & $10^{-2}$ s  \\ \hline
$\hat{T}_c$ & 10$^{-2}$~s & 10$^{-3}$~s & 10$^7$ s  \\ \hline
$\hat{\Lambda}_c$ & 10$^{-1}$~m & 10$^{-2}$~m & 10$^{19}$ m \\ \hline 
\end{tabular}
\caption{Typical order estimates from our model 
for gel, rubber, and slow earthquakes.
}
\end{table}

\subsection*{E. Order estimation of parameters}

We estimated the order of $\hat{\Lambda}_{c}$ and  $\hat{T}_{c}$ {\color{black} for soft matter or slow earthquakes.
By referring to the parameters relevant to laboratory experiments with soft matter like gels~\cite{Yamaguchi2011, Yamaguchi2016, Baumberger2002, Baumberger2006} or actual earthquakes~\cite{kano2010, Liu2005, Scholz1998},} we estimated the order of parameters listed in Table I using the following approximations:
\begin{align}
G-\beta =v_c^2 \beta,
\qquad
\Lambda_{c} \simeq \frac{c}{G-\beta}.
\end{align}
In addition, the following assumptions were applied {\color{black} 
for $(i)$ the soft matter experiments and 
$(ii)$ the tectonic plate:

$(i)$ In laboratory experiments, we indirectly obtained $\hat{D}_c$ from the other readings.
For Table I, $\hat{D}_c$ was adjusted to be close to the bifurcation point.

$(ii)$ The thickness $\hat{h}$ was chosen to satisfy
\begin{equation}
\hat{h} \simeq \hat{D}_c \frac{\hat{G}}{\hat{\sigma}(b-a)}  \simeq 10^{2} \mathrm{m},
\end{equation}
which justifies the thin layer elastic model with local coupling.
In addition, although $v_c$ can be estimated from $G$ and $\beta=b/a-1$ using Eq.~\eqref{ls12-2},
we instead suppose $v_c\simeq v_l$. 
}

The table leads to the conclusions that we could observe spatio-temporal chaos related to the present model in laboratory experiments~\cite{Yamaguchi2011,Yamaguchi2016, Baumberger2002, Baumberger2006}, but the order estimate does not show any quantitative agreement with results for geological phenomena.

{\color{black}

\subsection*{F. Derivation of the BK model }
The celebrated Burridge-Knopoff (BK) model~\cite{Burridge1967,Clancy2006,Thogersen2019a,Thogersen2021a} consists of an array of rigid blocks that are driven in the same direction by elastic springs attached with the driving plate. In addition, adjacent blocks are connected with another kind of springs. The BK model was originally proposed using rigid discrete elements~\cite{Burridge1967} and has been criticized as an essentially discrete model~\cite{Rice1993}. However, one can derive it by discretizing an elastic continuum subject to slow and long-wavelength deformation. Some attempts have been given in~\cite{Clancy2006,Thogersen2021a}. Here we derive the BK model in a slightly different manner so as to clarify the validity and limitations of the model.

We start from the analytical solution given in the literature~\cite{Rice1983}, which is the relation between the shear stress and the displacement on the interface between two elastic plates of finite thickness $\hat{h}$.
\begin{align}
	&\hat{\tau}^* (\hat{\omega},\hat{k})
	=
	-\hat{G}
	\frac{\sqrt{-(\hat{h}\hat{\omega})^2/\hat{c}^2+(\hat{h}\hat{k})^2}}{2\tanh{\left[\sqrt{-(\hat{h}\hat{\omega})^2/\hat{c}^2+(\hat{h}\hat{k})^2}\right]}}
	\hat{u}^*(\hat{\omega},\hat{k})
	\label{tau_gen}
	\\
	&\simeq
	-\frac{\hat{G}}{2\hat{h}}
	\left[
	1+\frac{1}{3}
	\left(-
	\frac{(\hat{h}\hat{\omega})^2}{\hat{c}^2}
	+
	(\hat{h}\hat{k})^2  \right)
	+\cdots
	\right]
	\hat{u}^*(\hat{\omega},\hat{k}),
	\label{tau_Taylor}
\end{align}
where $\hat{c}=\sqrt{\hat{G}/\hat{\rho}}$ denotes the elastic wave velocity.
Here we assume that $\hat{h}\hat{k} \ll 1$ and $\hat{h}\hat{\omega}/\hat{c} \ll 1$  and perform Taylor expansion with respect to these variables.
Note that the Fourier transform is performed from $\hat{t}$ and $\hat{x}$ to $\hat{\omega}$ and $\hat{k}$, respectively, and the superscript ${}^*$ indicates variables after each transformation.
We need to assume $\hat{u}^*(\hat{\omega},\hat{k})=0$ unless $\hat{h}\hat{k} \ll 1$ or $\hat{h}\hat{\omega}/\hat{c} \ll 1$.

Ignoring the higher-order terms in Eq. ~(\ref{tau_Taylor}) and going back to the real spacetime, we arrive at the local expression for the shear stress and displacement:
\begin{align}
	\hat{\tau} (\hat{t},\hat{x})
	&\simeq&
	-\frac{\hat{G}}{2\hat{h}}
	\left[
	1+\frac{\hat{h}^2}{3}
	\left(
	\frac{1}{\hat{c}^2} \frac{\partial^2}{\partial \hat{t}^2}
	-\frac{\partial^2}{\partial \hat{x}^2}
	\right)
	\right]
	\hat{u}(\hat{t},\hat{x})
\end{align}
Because this shear stress must balance with the friction, one obtains the following equation:
\begin{align}
	\frac{\hat{\rho}\hat{h}}{6}
	\frac{\partial^2}{\partial \hat{t}^2}\hat{u}(\hat{t},\hat{x})
	=
	-\frac{\hat{G}}{2\hat{h}}\hat{u}(\hat{t},\hat{x})
	+\hat{G}\frac{\hat{h}}{6}
	\frac{\partial^2}{\partial \hat{x}^2}\hat{u}(\hat{t},\hat{x})
	-{\rm friction}.
	\label{tan_thin}
\end{align}
This reduces to Eq.~(\ref{eqn02}) by changing the variables as follows: $\hat{u} \rightarrow \hat{u} -\hat{v}_l \hat{t}$, $4\hat{h} \rightarrow \hat{h}$, $2\sqrt{6}\hat{x} \rightarrow \hat{x}$, and $\hat{\rho}/6\rightarrow \hat{\rho}$.
}

\subsection*{G. Derivation of CGLE}
CGLE (Eq.~(\ref{TDCGL})) is derived from a model of a thin elastic layer that incorporates a rate-and-state friction law (Eqs.~(\ref{xi_time_evo})-(\ref{rs07})). Our derivation closely follows the scheme demonstrated in the reference \cite{Kuramoto1984}.

We first consider given equations without spatial coupling. 
Starting from following derivatives:
\begin{equation}
\frac{\partial F_{1}}{\partial \xi_1}= 1,  \nonumber
\end{equation}
\begin{equation}
\frac{\partial F_{2}}{\partial \xi_1}= -G,\,\,\,\, \frac{\partial F_{2}}{\partial \xi_2}=-\frac{1}{\xi_2+v_l},\,\,\,\, \frac{\partial F_{2}}{\partial \xi_3}=-1, \nonumber
\end{equation}
\begin{equation}
\frac{\partial^2 F_{2}}{\partial \xi_2^2}=\frac{1}{\left(\xi_2 + v_l\right)^2},\,\,\,\,\frac{\partial^3 F_{2}}{\partial \xi_2^3}=-\frac{2}{\left(\xi_2 + v_l\right)^3}, \nonumber
\end{equation}
\begin{equation}
\frac{\partial F_{3}}{\partial \xi_2}= -\xi_3-(1+\beta)\ln\left(\frac{\xi_2+v_l}{v_*}\right)-1-\beta, \nonumber
\end{equation}
\begin{equation}
\frac{\partial F_{3}}{\partial \xi_3}=-\xi_2-v_l, \,\,\,\, \frac{\partial^2 F_{3}}{\partial \xi_2 \partial \xi_3}=-1, \nonumber
\end{equation}
\begin{equation}
\frac{\partial^2 F_{3}}{\partial \xi_2^2}=-\frac{1+\beta}{\left(\xi_2 + v_l\right)},\,\,\,\frac{\partial ^3 F_{3}}{\partial \xi_2^3}=\frac{1+\beta}{\left(\xi_2 + v_l\right)^2}, \nonumber
\end{equation}
and the steady state solutions:
\begin{align}\label{swna10}
\xi_{1s} =-\frac{1}{G}\left\{\frac{\mu_{*}}{a} -\beta \ln\frac{v_l}{v_*}  \right\}, \nonumber \\
 \xi_{2s}=0,  \xi_{3s}=-(1+\beta)\ln\frac{v_l}{v_*},
\end{align}
the following expressions are obtained at the steady state:
\begin{equation} 
\left.\frac{\partial F_{1}}{\partial \xi_1}\right|_s= 1,  \nonumber
\end{equation}
\begin{align}
\left. \frac{\partial F_{2}}{\partial \xi_1}\right|_s= -G,\,\left. \frac{\partial F_{2}}{\partial \xi_2}\right|_s=-\frac{1}{v_l},\,\left.\frac{\partial F_{2}}{\partial \xi_3}\right|_s=-1, \nonumber \\
\left. \frac{\partial^2 F_{2}}{\partial \xi_2^2}\right|_s=\frac{1}{v_l^2},\,\left. \frac{\partial^3 F_{2}}{\partial \xi_2^3}\right|_s=-\frac{2}{v_l^3}, \nonumber
\end{align}
\begin{align}\label{swna13}
\left. \frac{\partial F_{3}}{\partial \xi_2}\right|_s= -1-\beta,\,\,\left. \frac{\partial F_{3}}{\partial \xi_3}\right|_s=-v_l,\,\,\left. \frac{\partial^2 F_{3}}{\partial \xi_2 \partial \xi_3}\right|_s=-1, \nonumber \\
\left. \frac{\partial^2 F_{3}}{\partial \xi_2^2}\right|_s=-\frac{1+\beta}{ v_l},\,\,\,\left. \frac{\partial ^3 F_{3}}{\partial \xi_2^3}\right|_s=\frac{1+\beta}{ v_l^2}. 
\end{align}
Loading velocity $v_l$ is set close to Hopf bifurcation point as $v_l=v_c\left(1+\varepsilon \right)$.
Expanding Eq.~\eqref{swna10}-\eqref{swna13}, by noting
\begin{align}\label{swna14}
\frac{1}{v_l}=\frac{1}{v_c}\left(1-\varepsilon \right),\,\, \frac{1}{v_l^2}=\frac{1}{v_c^2}\left(1-2\varepsilon \right), \nonumber \\
\frac{1}{v_l^3}=\frac{1}{v_c^3}\left(1-3\varepsilon \right),\,\, 
\end{align}
up to the first order of $\varepsilon$, we have
\begin{equation}
\left.\frac{\partial F_{1}}{\partial \xi_1}\right|_s= 1,  \nonumber
\end{equation}
\begin{align}
\left. \frac{\partial F_{2}}{\partial \xi_1}\right|_s= -G,\,\left. \frac{\partial F_{2}}{\partial \xi_2}\right|_s=-\frac{1}{v_c}\left(1-\varepsilon \right), \nonumber \\
\left.\frac{\partial F_{2}}{\partial \xi_3}\right|_s=-1,\,\, 
\left. \frac{\partial^2 F_{2}}{\partial \xi_2^2}\right|_s=\frac{1}{v_c^2}\left(1-2\varepsilon \right), \nonumber \\
\left. \frac{\partial^3 F_{2}}{\partial \xi_2^3}\right|_s=-\frac{2}{v_c^3}\left(1-3\varepsilon \right), \nonumber
\end{align}
\begin{align}\label{swna18}
\left. \frac{\partial F_{3}}{\partial \xi_2}\right|_s= -1-\beta,\,\,\left. \frac{\partial F_{3}}{\partial \xi_3}\right|_s=-v_c\left(1+\varepsilon \right), \nonumber \\
\left. \frac{\partial^2 F_{3}}{\partial \xi_2 \partial \xi_3}\right|_s=-1, \,\,\left. \frac{\partial^2 F_{3}}{\partial \xi_2^2}\right|_s=-\frac{1+\beta}{ v_c}\left(1-\varepsilon \right), \nonumber \\
\left. \frac{\partial ^3 F_{3}}{\partial \xi_2^3}\right|_s=\frac{1+\beta}{ v_c^2}\left(1-2\varepsilon \right).
\end{align}

The fluctuation around the steady state, $\delta\bm{\xi}$ obeys
\begin{equation}\label{swna20}
\frac{d \delta \xi_i}{d t}=({\bf L})_{ij}\xi_j+({\bf M}_0)_{ijk} \delta \xi_j \delta \xi_k + ({\bf N}_0)_{ijkl} \delta \xi_j \delta \xi_k \delta \xi_l,
\end{equation}
where
\begin{align}\label{swna21}
\left\{ 
\begin{array}{c}
({\bf L})_{ij}=({\bf L}_0)_{ij}+\varepsilon({\bf L}_1)_{ij} \\
({\bf M}_0)_{ijk}=\frac{1}{2}\left. \frac{\partial^2 F_i}{\partial \xi_j \partial \xi_k}\right|_s \\
({\bf N}_0)_{ijkl}=\frac{1}{6}\left. \frac{\partial^3 F_i}{\partial \xi_j \partial \xi_k \partial \xi_l}\right|_s
\end{array}
\right. .
\end{align}
The eigenvalue $\lambda$ is expanded according to the leading order of $\varepsilon$ as
\begin{equation}\label{swna22}
\lambda=\lambda_0+\varepsilon\lambda_1
\end{equation}
where
\begin{equation}\label{swna23}
\left\{ 
\begin{array}{c}
\lambda_0=\bm{\Xi}^* {\bf L}_0 \bm{\Xi} \\
\lambda_1=\bm{\Xi}^* {\bf L}_1 \bm{\Xi} 
\end{array}
\right.
\end{equation}
$\bm{\Xi}$ and $\bm{\Xi}^*$ are right and left eigenvector for ${\bf L}_0$, while
${\bf L}$ have following components: 
\begin{align}\label{swna24}
{\bf L}_0=
\left( 
\begin{array}{ccc}
0 & 1 &0 \\
-G & -\frac{1}{v_c} & -1 \\
0 & -1-\beta& -v_c
\end{array}
\right), \nonumber \\
{\bf L}_1=
\left( 
\begin{array}{ccc}
0 & 0 &0 \\
0 & \frac{1}{v_c} & 0 \\
0 & 0 & -v_c
\end{array}
\right).
\end{align}

The eigenvalues of ${\bf L}_0$ are $\alpha_c$ and $\pm i \omega_c$ where $\alpha_c = -1/v_c -v_c = -G/\sqrt{\beta (G-\beta)}$ and $\omega_c = \sqrt{G-\beta}$.
$\alpha_c <0$ indicates that the corresponding direction in the phase space is stable against perturbations. By contrast, $\pm i \omega_c$ have zero real part. Indeed, the eigenvalue space for $\pm i \omega_c$, become unstable at the Hopf bifurcation point. 
The right eigenvector $\bm{\Xi}$ corresponding for $i \omega_c$ is
\begin{equation}\label{swna28}
\bm{\Xi}=
\left( 
\begin{array}{c}
-i\frac{1}{\omega_c}  \\
1 \\
-\frac{\beta+1}{v_c+i\omega_c} 
\end{array}
\right)
=
\left( 
\begin{array}{c}
-i\frac{1}{\sqrt{G-\beta}}  \\
1 \\
\frac{-\sqrt{\beta}+i \beta}{\sqrt{G-\beta}} 
\end{array}
\right),
\end{equation}
while the left eigenvector $\bm{\Xi}^*$ for $i \omega_c$ is
\begin{equation}\label{swna30}
\bm{\Xi}^*=p\left(\frac{i G}{\sqrt{G-\beta }},1,\frac{i \sqrt{\beta} }{\left(\sqrt{\beta }-i\right)
	\sqrt{G-\beta }}\right),
\end{equation}
where $p$ is a normalization factor to satisfy $\bm{\Xi}^* \bm{\Xi}=1$ and 
\begin{equation}\label{swna31}
p =  \frac{\left(\sqrt{\beta }-i\right) (G-\beta)}{2 \left(G\sqrt{\beta
	} -\beta ^{3/2}-i G\right)}.
\end{equation}

\begin{figure}
	\begin{center}
		\includegraphics{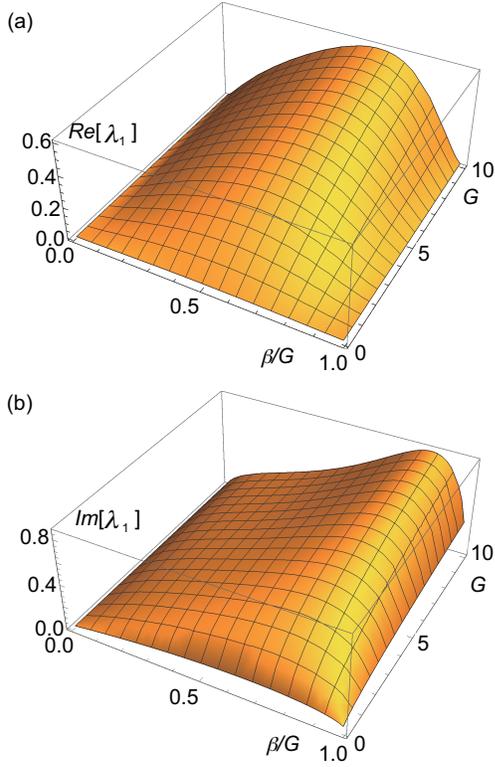}
		\caption{(a) $Re\left[\lambda_1 \right]$ and (b) $Im\left[\lambda_1 \right]$ with respect to various $G$ and $\beta/G$.}
		\label{lam1}
	\end{center}
\end{figure}

Using these expressions, one can obtain $\lambda_1$ as:
\begin{equation}\label{swna33}
\lambda_1 = \bm{\Xi}^* {\bf L}_1 \bm{\Xi}=\frac{\beta  \sqrt{G-\beta }}{-\beta ^{3/2}+\left(\sqrt{\beta }-i\right) G}.
\end{equation}
We can take the real and imaginary parts of $\lambda_1$, which are relevant for the analysis related with BF instability, as
\begin{align} 
Re\left[\lambda_1 \right] = \frac{(\beta  (G-\beta))^{3/2}}{\beta (G-\beta)^2+G^2}, \nonumber \\
Im\left[\lambda_1 \right]= \frac{\beta  G \sqrt{G-\beta }}{\beta (G-\beta)^2+G^2}. 
\label{swna34}
\end{align}
$Re\left[\lambda_1 \right]$ represents the growth rate of oscillation amplitude, and $Re\left[\lambda_1 \right] >0$ indicates that the steady state solution is not stable when $\varepsilon >0$.   As shown in Fig.~\ref{lam1}, $Re\left[\lambda_1 \right]$ is always positive in $G>0$, and $\beta>0$.

The higher order contribution of $\varepsilon$ is necessary to converge the amplitude of destablized oscillation, and is included in the coefficient $g$:
\begin{equation}\label{swna37}
g=-2\bm{\Xi}^{*}{\bf M}_{0}\bm{\Xi}\bm{Z}_0-2\bm{\Xi}^{*}{\bf M}_{0}\overline{\bm{\Xi}}\bm{Z}_{+}-3\bm{\Xi}^{*}{\bf N}_{0}\bm{\Xi}\bm{\Xi}\overline{\bm{\Xi}}, 
\end{equation}
where
\begin{align}\label{swna38}
\bm{Z}_{+}=-({\bf L}_0-2i \mathrm{I} \omega_c)^{-1}{\bf M}_{0}\bm{\Xi}\bm{\Xi},  \nonumber  \\
\bm{Z}_{0}=-2 {\bf L}_0^{-1}{\bf M}_{0}\bm{\Xi}\overline{\bm{\Xi}}.
\end{align}

For the calculation, one should use the following matrices to calculate $g$: 
\begin{widetext}
\begin{equation}
({\bf L}_0)^{-1}=
\frac{1}{G}\sqrt{\frac{\beta}{G-\beta}}
\left(
\begin{array}{ccc}
\beta & -\sqrt{\frac{G-\beta}{\beta}} & 1 \\
G\sqrt{\frac{G-\beta}{\beta}} & 0 & 0 \\
-G(1+\beta) & 0 & -G \\
\end{array}
\right), \nonumber
\end{equation}
\begin{equation}
{\bf L}_0-2 i \omega_c \mathrm{I}=
\left(
\begin{array}{ccc}
-2 i \sqrt{G-\beta } & 1 & 0 \\
-G & -2 i \sqrt{G-\beta }-\sqrt{\frac{\beta }{G-\beta }} & -1 \\
0 & -\beta -1 & -2 i \sqrt{G-\beta }-\sqrt{\frac{G-\beta }{\beta }} \\
\end{array}
\right), \nonumber
\end{equation}
\begin{align}
Det&=\mathrm{det} \left( {\bf L}_0-2 i \omega_c \mathrm{I} \right) =3\sqrt{G-\beta}\left\{\frac{G}{\sqrt{\beta}}+2 i (G-\beta) \right\}, \nonumber
\end{align}
and
\begin{align}
&\left({\bf L}_0-2 i \omega_c \mathrm{I}\right)^{-1} 
= \nonumber \\
&\frac{1}{Det} \left(
\begin{array}{ccc}
-4G +3\beta +2iG/\sqrt{\beta} & \sqrt{G-\beta}(1/\sqrt{\beta}+2i) & -1 \\
-G \sqrt{G-\beta}\left(1/ \sqrt{\beta }+2i\right) & 2 i(G-\beta ) \left(1/ \sqrt{\beta }+2i\right)  & -2i \sqrt{G-\beta} \\
G (\beta +1) & -2 i (\beta +1)\sqrt{G-\beta} &
4 \beta  -3 G +2 i \sqrt{\beta }
\end{array}
\right).\nonumber
\end{align}
\end{widetext}

For $\varepsilon$=0, 
\begin{equation}
\left. \frac{\partial^2 F_{2}}{\partial \xi_2^2}\right|_s=\frac{1}{v_c^2},\,\left. \frac{\partial^3 F_{2}}{\partial \xi_2^3}\right|_s=-\frac{2}{v_c^3}, \nonumber 
\end{equation}
\begin{align} 
\left. \frac{\partial^2 F_{3}}{\partial \xi_2 \partial \xi_3}\right|_s=-1,\,\left. \frac{\partial^2 F_{3}}{\partial \xi_2^2}\right|_s=-\frac{\beta+1}{v_c}, \nonumber \\
\left. \frac{\partial ^3 F_{3}}{\partial \xi_2^3}\right|_s=\frac{\beta+1}{ v_c^2}, \nonumber
\end{align}
and other second order and third order derivatives are zero.
From Eq.~(\ref{swna21}), we obtain,
\begin{align}
{\bf M}_0 \bm{\xi}^{(1)} \bm{\xi}^{(2)} \hspace*{16em}	 \nonumber \\
= \frac{1}{2}\left(
\begin{array}{c}
0 \\
\frac{1}{v_c^2}\xi_2^{(1)} \xi_2^{(2)} \\
-\xi_2^{(1)} \xi_3^{(2)} - \xi_3^{(1)} \xi_2^{(2)} -\frac{\beta+1}{v_c}\xi_2^{(1)} \xi_2^{(2)}
\end{array}
\right), \nonumber
\\
{\bf N}_0 \bm{ \xi}^{(1)} \bm{ \xi}^{(2)} \bm{ \xi}^{(3)}
=
\frac{1}{3v_c^3} \xi_2^{(1)}\xi_2^{(2)}\xi_2^{(3)}  \left(
\begin{array}{c}
0 \\
-1 \\
\frac{(1+\beta)v_c}{2}
\end{array}
\right). \nonumber
\end{align}
Inserting the components of the eigenvector $\bm{\Xi}$,
\begin{align} 
{\bf M}_0 \bm{\Xi}\bm{\Xi}
=\frac{1}{2}
\left(
\begin{array}{c}
0 \\
\frac{\beta }{ G-\beta } \\
- \left(\sqrt{\beta }+i\right)^2 \sqrt{\frac{\beta }{G -\beta}}
\end{array}
\right), \nonumber \\
{\bf M}_0 \bm{\Xi} \overline{\bm{\Xi}}=
\frac{1}{2}\left(
\begin{array}{c}
0 \\
\frac{\beta }{G-\beta } \\
(1-\beta ) \sqrt{\frac{\beta }{G-\beta}}
\end{array}
\right). \nonumber
\end{align}
Then, $\bm{Z}_{0}$ and $\bm{Z}_{+}$ leads,
\begin{equation} 
\bm{Z}_0
=\frac{\beta}{G(G-\beta)}
\left(
\begin{array}{c}
\beta \\
0 \\
G(1-\beta)
\end{array}
\right), \nonumber
\end{equation}
\begin{equation} 
\bm{Z}_{+}
=-\frac{1}{2 Det}
\left(
\begin{array}{c}
\frac{\beta (\sqrt{\beta} +4 i)}{\sqrt{G-\beta}} \\
2 i \beta (\sqrt{\beta}+4i) \\
\frac{i \sqrt{\beta} (\sqrt{\beta}+i) \left\{3G -8\beta + i \sqrt{\beta}(4\beta-3G)\right\}}{\sqrt{G-\beta}}
\end{array}
\right). \nonumber
\end{equation}

Thus, we obtain vector components for the following expressions as:
\begin{align}
{\bf M}_0 \bm{\Xi} \bm{Z}_0 =\left(
\begin{array}{c}
0 \\
0 \\
-\frac{(1-\beta ) \beta }{2 (G-\beta )}
\end{array}
\right), \nonumber
\end{align}
\begin{align} 
{\bf M}_0 \overline{\bm{\Xi}} \bm{Z}_{+} \hspace*{16em} \nonumber \\
=\left(
\begin{array}{c}
0 \\
-\frac{\left(\sqrt{\beta }+4 i\right) \beta ^{5/2}}{6 (G-\beta )^{3/2} \left\{2\sqrt{\beta} (G-\beta)-iG\right\}} \\
\frac{\beta \left[2\sqrt{\beta} (\beta^2 -2\beta +3G) + i\left\{ 10 \beta^2 -\beta(3G+8)+3G\right\} \right] }{12 (G-\beta) \left\{2\sqrt{\beta} (G-\beta)-iG\right\}}
\end{array}
\right), \nonumber
\end{align}
\begin{align} 
{\bf N}_0 \bm{\Xi} \bm{\Xi} \overline{\bm{\Xi}}
=
\left(
\begin{array}{c}
0 \\
-\frac{\beta^{3/2}}{3 \left(G-\beta \right)^{3/2}} \\
\frac{\beta  (\beta +1)}{6
	(G-\beta )}
\end{array}
\right). \nonumber
\end{align}
Collecting these resulting expressions,
\begin{align} 
-2 \bm{\Xi}^{*} {\bf M}_0 \bm{\Xi} \bm{Z}_0= -\frac{i (\beta -1) \beta ^{3/2}}{2\sqrt{G-\beta
	} \left\{2\sqrt{\beta} (G-\beta)-iG\right\} }, \nonumber
\end{align}
\begin{align}
&-2\bm{\Xi}^{*}{\bf M}_0 \overline{\bm{\Xi}} \bm{Z}_{+}= \hspace*{10em} \nonumber \\
&\frac{\left(1-i \sqrt{\beta }\right) \beta ^{3/2} \left(10 i \beta ^{3/2}+2 \beta ^2-3 i
	\sqrt{\beta } G+3 G\right)}{12 \sqrt{G-\beta }\left\{2\sqrt{\beta} (G-\beta)-iG\right\} \left\{\sqrt{\beta}(G-\beta)-iG\right\}}, \nonumber
\end{align}
\begin{align}
-3\bm{\Xi}^{*} {\bf N}_0 \bm{\Xi} \bm{\Xi} \overline{\bm{\Xi}}=
\frac{i \beta^{3/2}  \left(\beta +2 i \sqrt{\beta }+3\right) }{4\sqrt{G-\beta}
	\left\{\sqrt{\beta}(G-\beta)-iG\right\}}, \nonumber
\end{align}
and finally,
\begin{align}\label{swna56}
g=\frac{-i \beta ^2 (\beta +1) (9 G-8\beta)}{6 \sqrt{G-\beta } \left(2 \beta  ( G- \beta)^2-3
	i \sqrt{\beta } G (G-\beta)-G^2\right)}. 
\end{align}

$g$ has the following real and imaginary parts:
\begin{widetext}
\begin{align}\label{swna57}
Re\left[g \right] =\frac{\beta ^{5/2} (\beta +1) G \left(8 \beta ^2-17 \beta  G+9 G^2\right)}{2 \sqrt{G-\beta }
	\left(4 \beta ^6-16 \beta ^5 G+24 \beta ^4 G^2+\beta ^3 (5-16 G) G^2+2 \beta ^2 G^3 (2
	G-5)+5 \beta  G^4+G^4\right)},
\end{align}
\begin{align}\label{swna58}
 Im\left[g \right] 
=\frac{\beta ^2 (\beta +1) \left(16 \beta ^4-50 \beta ^3 G+52 \beta ^2 G^2-2 \beta  G^2 (9
	G+4)+9 G^3\right)}{6 \sqrt{G-\beta } \left(4 \beta ^6-16 \beta ^5 G+24 \beta ^4 G^2+\beta
	^3 (5-16 G) G^2+2 \beta ^2 G^3 (2 G-5)+5 \beta  G^4+G^4\right)}.
\end{align}
\end{widetext}
From $\lambda_1$ and $g$, the oscillation amplitude, $R_s$,  is given as:
\begin{align}\label{swna61}
R_s=\sqrt{\frac{Re[\lambda_1]}{Re[g]}}=\sqrt{2} \sqrt{\frac{(G-\beta) \left(4 \beta (G-\beta)^2+G^2\right)}{\beta
		(\beta +1) G ( G+8( G- \beta))}}.
\end{align}
In addition, a part of the expression included in the condition of BF instability (Eq.~\eqref{BF_condition}) is given as:
\begin{equation}\label{swna59}
\frac{Im\left[g \right]}{Re\left[g \right]}=\frac{G^2 - 2 \beta(G-\beta)^2}{3 \sqrt{\beta } G (G -\beta)}.
\end{equation}

Based on the analyses on the spatially homogeneous situation, we then include the local diffusional coupling. The original equations to be considered are Eqs.~\eqref{xi_time_evo}-\eqref{rs07}, but 
the auxiliary viscose friction $\nu\partial^2 \xi_2 / \partial x^2$ is additionally included in $d \xi_2/dt$ to estimate the impact on the dynamics.
The complex diffusion constant is
\begin{equation}\label{}
d = \bm{\Xi}^* \mathrm{D} \bm{\Xi}, \quad \text{where} \quad \mathrm{D}=
\left(
\begin{array}{ccc}
0&0&0  \\
c^2&\nu&0 \\
0&0&0
\end{array}
\right). \nonumber
\end{equation}

From eigenvectors shown in Eqs.~\eqref{swna28}, \eqref{swna30}, 
\begin{equation}\label{}
d=\frac{c^2 \left(i\sqrt{\beta }+1 \right) \sqrt{G-\beta } -\nu (G-\beta)\left(\sqrt{\beta }-i\right)   }{2 \left[   \beta^{3/2}- \left(\sqrt{\beta }-i \right)G\right]},
\end{equation}
where the real part and imaginary parts are
\begin{align}\label{}
&Re\left[d \right] \nonumber \\
&=\frac{c^2 \beta^{3/2} \sqrt{G-\beta} + \nu (G-\beta) \left[G+\beta (G-\beta) \right] }{2 \left[\beta(G-\beta)^2+G^2   \right]}, \nonumber
\end{align}
\begin{align}\label{}
&Im\left[d \right] \nonumber \\
&=\frac{\sqrt{G-\beta } \left[c^2 \left\{\beta^2 -(\beta+1)G  \right\}  +\nu\beta^{3/2}\sqrt{G-\beta} \right]}{2 \left[ \beta(G-\beta)^2+G^2   \right]}, \nonumber
\end{align}
and
\begin{align}\label{}
&\frac{Im\left[d \right]}{Re\left[d \right]}= \frac{c^2 \left\{\beta^2 -(\beta+1)G  \right\} +\nu  \beta^{3/2} \sqrt{G-\beta } }{c^2 \beta^{3/2} +\nu \sqrt{G-\beta } \left[G+ \beta(G-\beta)\right] }.
\end{align}
With $\nu=0$, the expressions are simplified as:
\begin{align}
d= \frac{c^2 \left(i\sqrt{\beta }+1 \right) \sqrt{G-\beta }  }{2 \left[   \beta^{3/2}- \left(\sqrt{\beta }-i \right)G\right]}, \nonumber \\
Re\left[d \right]=\frac{c^2 \beta ^{3/2}\sqrt{ (G-\beta )}}{2 \left[\beta(G-\beta)^2+G^2   \right]}, \nonumber \\
Im\left[d \right]=\frac{c^2 \sqrt{G-\beta } \left[\beta ^2-(\beta+1)  G\right]}{2 \left[\beta(G-\beta)^2+G^2   \right]}, \nonumber \\
\frac{Im\left[d \right]}{Re\left[d \right]}=
\frac{\beta ^2-(\beta +1) G}{\beta ^{3/2}}. \label{seqdratio}
\end{align}

The condition for the Benjamin-Feir instability with $\nu=0$ is
\begin{align}\label{}
1+\frac{Im\left[d \right]}{Re\left[d \right]} \cdot \frac{Im\left[g \right]}{Re\left[g \right]}= \hspace*{10em} \nonumber \\
\frac{\left[2 \beta ( G-\beta) - G\right] \left[\beta (G-\beta)^2+G^2 \right]}{3 \beta^2  G (G-\beta)}<0.
\end{align}
From $G>0, \beta>0$ and $G-\beta>0$, the condition for Benjamin-Feir instability is
\begin{equation}\label{}
G>2\beta (G-\beta), \nonumber
\end{equation}
or equivalently,
\begin{equation}\label{}
G<\frac{1}{2 \frac{\beta}{G}\left(1-\frac{\beta}{G}\right)}.
\end{equation}

\bibliographystyle{apsrev4-1}

\end{document}